# A dynamic logic method for determining behaviors of biological networks

SuPing Lyu

*Medtronic Corporate Science and Technology, 710 Medtronic Parkway, Minneapolis, MN 55432*

*(June 10, 2008)*

**ABSTRACT** A dynamic logic method was developed to analyze molecular networks of cells by combining Kauffman and Thomas's logic operations with molecular interaction parameters. The logic operations characterize the discrete interactions between biological components. The interaction parameters (e.g. response times) describe the quantitative kinetics. The combination of the two quantitatively characterizes the discrete biological interactions. A number of simple networks were analyzed. The main results include: we proved the theorems to determine bistable states and oscillation behaviors of networks, we showed that time delays are essential for oscillation structures, we proved that single variable networks do not have chaotic behaviors, and we explained why one signal can have multiply responses. In addition, we applied the present method to the analysis of the MAPK cascade, feed-forward loops, and mitosis cycle of budding yeast cells.

**Keywords:** molecular network, logic, signal transduction, bistable, oscillation, and systems biology

## 1. INTRODUCTION

A large number of nucleic acids, proteins, and other molecules have been discovered in biological systems. These molecules interact with each other and form complicated and large scale molecular networks [1-6]. This has driven research toward systems levels. However, due to the complexity of large systems, it is challenging to elucidate their structures, functions, interactions, and dynamic behaviors based on experimental approaches alone. Statistical analysis, quantitative modeling, and logic analysis have been pursued to understand the complexity. Statistical analysis of these networks successfully reveals a lot of interesting features [7-20] such as motifs [7-8] (feedback or feedforward control loops formed by two to four proteins), modules [10-12], small-world properties [13-15] (shortest path between protein pairs is small), power law distribution of interactions [16-18], and centrality [19-20] (a few highly connected proteins play central role in mediating interactions among numerous less connected proteins). However, statistical studies have not focused enough on quantitative dynamics of molecular networks.

Quantitative modeling has been conducted based on deterministic process for the purpose of understanding and predicting the dynamic properties of the networks [21- 46]. Among these properties, there are robustness [21, 26, 46], noise resistance [29, 32], sensitivity [30, 33], bistability and oscillation [22, 26, 28, 42, 45, 46]. Excellent review can be found in the recent book by Alon [46] and reference 41. Stochastic methods (e.g. Stochastic Petri Net) have also been developed and have been used to extract kinetic parameters by reverse-engineering [47- 49]. Although quantitative modeling has very successful in understanding a number of important structures and interactions of molecular networks of cells, use of these methods has been limited by the lack of quantitative kinetic parameters that are needed to quantify the models [41]. This gap between the need and availability of quantitative parameters can become greater due to the explosive discovery of new proteins and nucleic acids without sufficient quantification.

This un-met need in quantitative parameters may lead to a question: is it practically possible and needed for experimental biologists to measure the kinetic parameters for all the biological components discovered? Even if all those parameters could be measured, would it be feasible to solve such a large number of differential equations and interpret the results clearly? This problem is similar to that in thermodynamics where it is neither possible nor necessary to solve large numbers of differential equations of all the molecules of the systems. Instead, the statistical mechanics has been proved valuable. As such, a fundamental question arises: how quantitative should the research of biological networks be? Or in other words, what does the concept of "quantitativeness" mean to biology?

The biology research itself may help to answer this question. Biologists commonly use binary descriptions such as activation and deactivation, phosphorylation and dephosphorylation, and methylation and demrethylation all the time. It seems biological systems can be well described with such qualitative words except that these descriptions do not provide us kinetic information such as response times and dose effects. Recently, there has been an increasing amount of evidence that shows the components of molecular networks of cells take ON or *OFF* states (all active or none) [21-28, 30, 32, 42-46 50-63]. Further more it has been proposed that the functions of network are inherent properties of the network itself, rather than the fine-tuning properties of individual network components [21].

The above questions about quantitativeness of biology suggests that discrete mathematic tools may be suitable for these systems. In deed, logic and graph theories, both are discrete, have been used to study biological systems and the efforts were fruitful [41, 64-82]. For example, Thomas used truth tables of discrete logic variables to analyze molecular networks of cells. It was found molecular networks can be stable or oscillate based on their feedback structures [66]. Such discrete steady state behaviors seem to be very common. However, to treat the biological components with discrete variables may be criticized for a lack of quantitative kinetic information such as rates, durations, and doses. Therefore, some necessary kinetic parameters such as response times and dose effects need to be incorporated into discrete methods. Also, the existing logic method depends on truth tables to determine the steady state solutions of systems [66]. Complexity of calculation increases quickly as systems become large. Therefore, a non-truth table approach is needed to extend the logic methods.

There is another question related to the dynamic nature of the interactions among network components. It has been generally accepted that genes, RNAs, proteins, and small molecules are generated or activated when needed, but all are degraded or deactivated after their jobs are completed. The reason behind this is that the degradation reactions allow systems to return to their pre-stimulation states and become ready for the next activation/deactivation cycles. If there is no degradation or deactivation, the components will stay activated and the systems become static. For this reason, auto-degradation is often included in modeling studies of activation interactions [32, 34-36, 38, 39, 83]. However, in real situations, deactivation interactions occur as often as that of activation interactions. The above auto-degradation processes need to be generalized in order to treat both the activation and deactivation interactions to ensure that molecular networks can always respond to stimulations to keep the network systems biologically active.

In the present paper, we extended the idea that biological networks are discrete to a new one that biological networks are dynamic and discrete. We incorporated logic methods with dynamic parameters and developed a dynamic logic method. We first summarized the above questions into two assumptions, the discrete state assumption and readiness assumption. We proposed a method that incorporates discrete logic-graph theory with quantitative response times. Then, we develop algorithm with the simplest network systems and generalized it to large systems. We demonstrated how the method could be used to predict the states of networks (e.g. bistable or oscillatory), calculate the kinetic parameters (e.g. oscillation periods), and explain the experimental and modeling results in literature.

## 2. NOMENCLATURE

Molecular networks of cell have been described with the terminology of graph theory because they are analogous to directed signed graphs [41, 65]. The components of networks (protein, Gene, mRNA, or metabolic compounds) have been called nodes (represented with capital letters, *A*, *B*, etc.). Nodes can be at active states (*A*, *B*…) or inactive states ( $\overline{A}$ , $\overline{B}$ …). Node states are also called node values. The interactions among nodes called edges. Edges have directions and can be described with logic expression. For example, a logic structure $A \to B$ is an edge in which node *A* activates node *B*. $A-| B$ is an edge in which node *A* deactivates *B*. We refer to the nodes that stimulate other nodes as "stimulators" and the nodes that respond to stimulations as "responders." Before stimulations, the states of the responders are called "ready states." After the nodes fully responding to stimulation, their states are referred to as "responding states." A node can have one or more stimulators as well as one or more responders. A node does not have any stimulator is referred to as root. A node does not have any responder is referred to as leaf. Depending on the numbers of stimulators and responders of their nodes, systems can be linear chains, cycles, trees, or complicated networks with trees and cycles embedded in.

Response time of an interaction (edge) is referred to as the interval from the time when the stimulator starts to act on its responder to the time when the responder reaches its responding state. For example, $t_{A \to B}$ is the response time for *A* to activate *B* and $t_{\overline{A} \to \overline{B}}$ is the response time for $\overline{A}$ to deactivate *B* (becoming $\overline{B}$ ). The response times are an integral representative of the node interaction kinetics. A node may be at its active state for a certain period of time then changes to its inactive state. We refer the time within which a node is active as to active pulse width ($T_A$). Inactive pulse width ($T_{\overline{A}}$) is defined similarly.

## 3. ASSUMPTIONS

**Discrete State Assumption**. Network components (nodes) take discrete states such as active states and inactive states. Interactions (edges) among the nodes are also discrete such as activation and deactivation.

**Readiness Assumption.** A component will automatically return to its ready state from responding state within a response time after the stimulation stops. The ready states can be either active or inactive (corresponding to deactivation or activation). This generalizes the auto-degradation processes discussed in Introduction section. This transition guarantees the components of the network will be ready for the next stimulation. This readiness assumption defines the relaxation process of a system. This property may be understood in term of the evolutionary processes that lead biological systems to fit their environments [51]. It was argued that the biological functions follow the structures [84]. Then, the biological systems must have mechanisms to ensure their abilities to respond to stimulations. Examples of such mechanisms may include reversible reactions (phosphorylation and dephosphorylation) and combined synthesis and degradation (or dilution).

## 4. DYNAMIC LOGIC REPRESENTATIONS

### 4.1 Truth table representation

Before presenting the dynamic logic representation, we like to briefly review some current logic methods based on truth tables. Truth tables are lists of all the values that logic variables can take. Logic relationships can be determined based on those lists. Thomas et al. analyzed biologic networks using truth tables as the basic tool [66]. However, the truth tables needed for biological analysis are significantly different from that in mathematic logic and electric circuitry. We illustrate this difference with a simple logical example $A \rightarrow B$ (Figure 1).

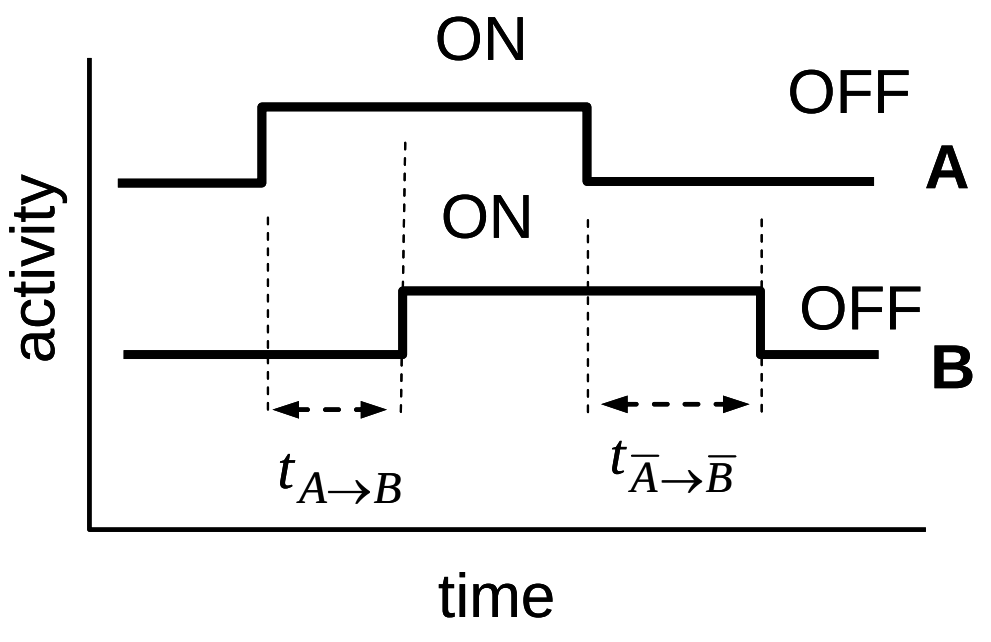

Figure 1. The states and interactions of molecular network *A*→ *B*. Both nodes *A* and *B* are assumed to have only *ON* and *OFF* states. Therefore, their activity curves have step shapes. Similar ideas are discussed in reference 66.

In mathematical logic, $A \rightarrow B$ means if variable *A* is true, variable *B* is true. If A is not true ( $\overline{A}$ ), *B* can be either true (*B*) or not true ( $\overline{B}$ ), so *B* is undetermined. The true table includes these three cases is described in Table 1.

Table 1. Truth table of $A \rightarrow B$ based on mathematic logic and electric circuitry.

| Mathematic logic | | Electric circuitry | |
|---|---|---|---|
| $A$ | $B$ | $A$ | $B$ |
| $\overline{A}$ | $B$ | $\overline{A}$ | $\overline{B}$ |
| $\overline{A}$ | $\overline{B}$ | | |

In electric circuitry, a relationship $A \rightarrow B$ means $B = A$. That is if A is true $B$ is true and if $A$ is not true ($\overline{A}$) $B$ is not true ($\overline{B}$) (Table 1). The values or states of those variables are determined, but there is no time delay between them.

Table 2. Truth tables of logic variables and functions for a biological system $A \rightarrow B$.

| | *External activation* | *Variables* | | *Functions* | |
|---|---|---|---|---|---|
| 1 | *Off* | $\overline{A}$ | $\overline{B}$ | $\overline{a}$ | $\overline{b}$ |
| 2 | *On* | $\overline{B}$ | $\overline{B}$ | $a$ | $\overline{b}$ |
| 3 | *On* | $A$ | $\overline{B}$ | $a$ | $b$ |
| 4 | *On* | $A$ | $B$ | $a$ | $b$ |
| 5 | *Off* | $A$ | $B$ | $\overline{a}$ | $b$ |
| 6 | *Off* | $\overline{A}$ | $B$ | $\overline{a}$ | $\overline{b}$ |
| 7 | *Off* | $\overline{A}$ | $\overline{B}$ | $\overline{a}$ | $\overline{b}$ |

In biological networks, $A \rightarrow B$ has a similar meaning to that of electric circuitry, but there is a time delay. Assume both *A* and *B* are inactive at beginning. When *A* becomes active, its action to activate *B* starts immediately. But, *B* become active only after the activation action of *A* continues for a response time $t_{A \rightarrow B}$. This is the so called separation of logic variables and functions (actions). In Thomas' approach, two sets of logic variables were used to represent the variables (*A* and *B*) and their functions (*a* and *b*). Then two truth tables were used. For example, this $A \rightarrow B$ relationships can be described in Table 2. At beginning, both nodes are inactive ($\overline{A}$ and $\overline{B}$). There is no external signal to activate them, so they tend to remain inactive. For such a case, the function values are $\overline{a}$ and $\overline{b}$. At certain time, an external signal activates *A* and keep it at active state. *A* immediately starts to activate $\overline{B}$. So the values of functions are *a* and *b* (*row 3*). After an activation response time, $\overline{B}$ becomes active (*B*) (*row 4*). At this moment, the system reaches its steady state responding to the external activation. After another time, the external signal disappears. Deactivation action on $A$ starts and function values are $\bar{a}$ and $b$ (*row 5*). Subsequently, $A$ is deactivated (*row 6*) and then $B$ is deactivated (row 7). Then the system reaches its inactive steady state ( $\bar{A}$ and $\bar{B}$ ). Therefore, the steady state values of biological system are similar to that of electric circuitry. However, the change from one state to the other is delayed by response times. Such delays are key dynamic characteristics of biological systems. However, in this true table approach, the response times are not effectively incorporated into its representation. We cannot study both the logic relationships and dynamics (response times) in the same time. Also, truth tables can get very large for complicated systems.

### 4.2 Dynamic logic sequence

In the present paper, we proposed a different symbolic system that combines logic arguments and response times (interactions kinetics) into a dynamic logic sequence. This sequence can represent the logic relationships and delay times of a biological network in a compact form. Further more, it allows the analysis of networks to be done in a general way instead of case by case as with the truth table methods. We introduce this method by using the same $A \rightarrow B$ example of Figure 1.

A dynamic logic sequence of a system is defined as a sequence composed of stimulator state, the response time, and the responder state. For the system $A \rightarrow B$, one dynamic logic sequence is $A_{t_{A \rightarrow B}} B$, representing that $A$ activates $B$ within a response time $t_{A \rightarrow B}$. Similarly, $\bar{A}_{t_{\bar{A} \rightarrow \bar{B}}} \bar{B}$ represents that $\bar{A}$ deactivates $B$ to $\bar{B}$ within a response time $t_{\bar{A} \rightarrow \bar{B}}$. Thus, the logic relationships and response times (kinetics) are combined into a single form that is compact compared to the truth table (Table 2). For a general network, its dynamics logic sequences contain the states of all its nodes and the corresponding response times in an order defined by its logic structure. It can be determined based on the following rules:

R1. A dynamic logic sequence of a network contains maximum number of its node states.

R2. At any given time within a sequence, a node takes only one state (e.g. either $A$ or $\bar{A}$ ). However, a node can change its state over time due to changes of its stimulators and it can appear in a sequence for more than once.

R3. The note states in a sequence meet logic relationships. Stimulator state is earlier than its responder state. The time distances between stimulator states and the responder state are equal to the response times. Earlier event is written down in the left side of a sequence. Only root is allowed at the beginning of a sequence. And only a leaf is allowed at the end of a sequence.

R4. If a network system has more than one sequence, these sequences are written down in an addition format (e.g. $A_{t_{A \rightarrow B}} B + \bar{A}_{t_{\bar{A} \rightarrow \bar{B}}} \bar{B}$ for $A \rightarrow B$).

From *R1* and *R3*, if a node state of one sequence is the condition for a node state of another sequence to change, these two sequences should be combined into a more complete sequence. The most complete sequences are referred to as the values of a network. For example, sequences $A_{t_{A \to B}} B$ and $\overline{A}_{t_{\overline{A} \to \overline{B}}} \overline{B}$ are the two values of system $A \to B$.

A node state with its response time in immediate front of it is referred to as the action condition of the node. For example $t_{A \to B} B$ is the action condition of *B*. It indicates that *B* becomes active if its activator (*A*) was activated $t_{A \to B}$ ago. A node state with its response time immediately after it is referred to as a logic equivalent of its responder. For example, $A_{t_{A \to B}}$ is a logic equivalent of *B*, or $A_{t_{A \to B}} \leftrightarrow B$. It indicates that node *B* becomes active is logically equivalent to that node *A* becomes active $t_{A \to B}$ ago.

### 4.3 Time sequence

The dynamic logic sequence allows us to write down the time sequences of the events that occur within a biological network. For $A \to B$, we have the time sequence of node *A* as

$$\overline{A}_{(t_0)} A_{(t_A)} \overline{A} \quad (1)$$

That is *A* is inactive before instant $t_0$. It changes to active state at instant $t_0$ and becomes inactive again at instant $t_A$. The active pulse width of node *A* is $T_A = t_A - t_0$. The time sequence of node *B* is

$$\overline{B}_{(t_0 + t_{A \to B})} B_{(t_A + t_{\overline{A} \to \overline{B}})} \overline{B} \quad (2)$$

The whole networks have time sequence as well:

$$\overline{A}\overline{B}_{(t_0)} A\overline{B}_{(t_0 + t_{A \to B})} AB_{(t_A)} \overline{A}B_{(t_A + t_{\overline{A} \to \overline{B}})} \overline{A}\overline{B} \quad (3)$$

Equation 3 is a 1-D representation of the two sets of truth tables of Thomas' method (Table 2) with delay times being incorporated. This sequence contains the node states, node state changes, and the change times. Therefore, this sequence provides all the information about the dynamic process of networks. In electric circuitry, only AB and $\overline{A}\overline{B}$ are allowed. But, in biological biology, $A\overline{B}$ and $\overline{A}B$ are also allowed. The time sequence of network can be simplified by including only the nodes that have state changes. For example, the above time sequence (Equation 3) can simplified into

$$\overline{A}\overline{B}_{(t_0)} A_{(t_0 + t_{A \to B})} B_{(t_A)} \overline{A}_{(t_A + t_{\overline{A} \to \overline{B}})} \overline{B} \quad (4)$$

It should be noted that times in dynamic logic sequences are response times (how long). The times in time sequences are instants. To distinguish this, instants in time sequences are in brackets.

## 5. ALGORITHM

The procedure to calculate dynamic logic sequences and time sequences of networks depends on the specific network structures. The simplest network systems are those that have two components and one interaction relationship (two nodes and one edge). Network systems can get complicated by having more nodes. However, systems are still simple as long as each node has no more than one stimulator. We call this type of systems as single variable network systems. When there are two or more stimulators for some nodes, the interactions within the systems become complicated. These systems are referred to as multiple variable network systems. We will discuss the calculations of values of these three types of networks separately.

### 5.1 Simple network systems with two components and single interaction

One of the simplest system is $A \to B$. We have already known that this system have two values (dynamic logic sequences), $A_{t_{A\to B}}B$ and $\overline{A}_{t_{\overline{A}\to\overline{B}}}\overline{B}$. Why these are the only two values? The first value is based on the definition of this system that is if *A* is active *B* becomes active after $t_{A\to B}$. Based on mathematical logic, if *A* is not active, *B* can be either *B* or $\overline{B}$. So, there might be another two values. However, the readiness assumption says that the component of networks, after its stimulation signal turns off, will return to its ready state. for this case, the ready state of *B* is $\overline{B}$. Therefore, if *A* becomes inactive, *B* will return its inactive state $\overline{B}$. Then, we have only one more value, $\overline{A}_{t_{\overline{A}\to\overline{B}}}\overline{B}$, for system $A \to B$. This value defines the relaxation process. The complete values of this network, $V(A \to B)$, are

$$V(A \to B) = A_{t_{A\to B}}B + \overline{A}_{t_{\overline{A}\to\overline{B}}}\overline{B} \tag{5}$$

Where, "+"means the system can take either one of the two values depending on the external conditions. This example demonstrated that the readiness assumption allows us to define the network dynamics without additional information about its relaxation direction.

A second simple network is a system composed of *A* and *B* with a deactivation relationship, $A\!-\!|\,B$. The value of this system can be obtained in a similar way.

$$V(A\!-\!|B) = A_{t_{A\to\overline{B}}}\overline{B} + \overline{A}_{t_{\overline{A}\to B}}B \tag{6}$$

The time sequence of this system is

$$\overline{A}B_{(t_0)}AB_{(t_0+t_{A\to\overline{B}})}A\overline{B}_{(t_A)}\overline{A}\overline{B}_{(t_A+t_{\overline{A}\to B})}\overline{A}B \tag{7}$$

We can confirm that the logic relationship "$A\!-\!|\,B$" is equivalent to "$\overline{A} \to B$" by simply comparing the values of Equations 5 and 6. That is,

$$(A\!-\!|\,B) \leftrightarrow (\overline{A} \to B) \tag{8}$$

This is a very useful relationship to convert the deactivation logic forms into activation forms.

### 5.2 Single-variable network systems with more one edge

Networks with many nodes and edges can be considered as combinations of simple activation or deactivation interactions ($A \to B$ or $A\!-\!|\,B$). Therefore, the values of complicated networks can be calculated by combining the values of all the individual edges of systems. The procedure is as the follows.

S1. Calculate the values of each individual edge. If there is more than one value, write them down in an addition format (*Rule 4*).

S2. Combine the values of edges that have common nodes. The sequential orders of the different edges in the combinations need to satisfy the requirements defined by *Rule 3*, that is, the stimulators are in the left side of responders. Repeat this with all the nodes (*Rule 1*).

S3. At any given place in sequences, a node takes only one value. If a node has different values at the same time, this sequence is invalid according to *Rule 2* and needs to be rejected.

S4. If there are more than one dynamic logic sequence, and the ending node state of one sequence is the same as the starting node state of another sequence, combine these two sequences together *(Rule 3).*

We use an example network, $A \to B \to C$ to demonstrate this procedure. The sequential order of the edges in this network is $A \to B$ then $B \to C$. The combination should take the same order. Taking all the different combinations of the values of these two edges, we have

$$
\begin{aligned}
&V(A \to B \to C) = V(A \to B)V(B \to C) \\
&= (A_{t_{A\to B}} B + \overline{A}_{t_{\overline{A}\to\overline{B}}} \overline{B})(B_{t_{B\to C}} C + B_{t_{\overline{B}\to\overline{C}}} \overline{C}) \\
&= A_{t_{A\to B}} BB_{t_{B\to C}} C + A_{t_{A\to B}} B\overline{B}_{t_{\overline{B}\to\overline{C}}} \overline{C} + \overline{A}_{t_{\overline{A}\to\overline{B}}} \overline{B}B_{t_{B\to C}} C + \overline{A}_{t_{\overline{A}\to\overline{B}}} \overline{B}\overline{B}_{t_{\overline{B}\to\overline{C}}} \overline{C}
\end{aligned}
$$

According to Sequence *Rule 2*, each node of the systems can take only one value at any given time. The value of the node *B* in the second and third items of the above combinations has different values at same times. Therefore, these two items are rejected. In the first and fourth items, both *A* and *B* have one value at any given time. Therefore, first and fourth items are the valid values of this network. Merging *BB* and $\overline{B}\overline{B}$ into *B* and $\overline{B}$, we have

$$
V(A \to B \to C) = A_{t_{A\to B}} B_{t_{B\to C}} C + \overline{A}_{t_{\overline{A}\to\overline{B}}} \overline{B}_{t_{\overline{B}\to\overline{C}}} \overline{C} \tag{9}
$$

The delay time from A to C is $t_{A\to B} + t_{B\to C}$ and $\overline{A}$ to $\overline{C}$ is $t_{\overline{A}\to\overline{B}} + t_{\overline{B}\to\overline{C}}$.

In the above examples, the flow of stimulation is linear. Now, we consider a system in which *A* and *B* mutually deactivate each other such that $A \vdash\!\dashv B$. The values of this system is the combination of that of $A -| B$ and $B -| A$, or

$$
\begin{aligned}
&(A_{t_{A\to\overline{B}}} \overline{B} + \overline{A}_{t_{\overline{A}\to B}} B)(B_{t_{B\to\overline{A}}} \overline{A} + \overline{B}_{t_{\overline{B}\to A}} A) \\
&= A_{t_{A\to\overline{B}}} \overline{B}B_{t_{B\to\overline{A}}} \overline{A} + A_{t_{A\to\overline{B}}} \overline{B}\overline{B}_{t_{\overline{B}\to A}} A + \overline{A}_{t_{\overline{A}\to B}} BB_{t_{B\to\overline{A}}} \overline{A} + \overline{A}_{t_{\overline{A}\to B}} B\overline{B}_{t_{\overline{B}\to A}} A
\end{aligned}
$$

The first and fourth items are not valid solutions since the component *B* takes different values at the same time which violates the *Rule 2*. The second and the third items are valid. Then we have

$$
V(A \vdash\!\dashv B) = A_{t_{A\to\overline{B}}} \overline{B}_{t_{\overline{B}\to A}} A + \overline{A}_{t_{\overline{A}\to B}} B_{t_{B\to\overline{A}}} \overline{A} \tag{10}
$$

The two values of this double deactivation system are similar to that of the single deactivation system (Equation 6). However, the double deactivation system is stable. For example, when the system is at state $A_{t_{A\to \bar{B}}}\bar{B}_{t_{\bar{B}\to A}}A$, fluctuation of *A* to $\bar{A}$ is inhibited by $\bar{B}$ , and fluctuation of $\bar{B}$ to *B* is inhibited by *A*. The system stays at $A\bar{B}$ . The same is true for the other state $\bar{A}_{t_{\bar{A}}}B_{t_B}\bar{A}$ . But, for the single deactivation system (Equation 6), *B* does not have an inhibition effect on *A*. Therefore, it is not stable. It should be noticed that although a double deactivation system can have two stable states, switching from one state to the other needs to be driven by external conditions of the system. The system may stay with single state if external condition does not drive it to change. This can be demonstrated using the time sequence of the system.

Assuming the system is $\bar{A}B$ . An external signal comes in and turns on *A* at time zero. Based on Equation 10, *B* will be turn off at $t_{A\to\bar{B}}$. Then, after $t_{\bar{B}\to A}$, $\bar{B}$ starts to stabilize *A* at *ON* state. At this moment, even the external signal goes away, the system remains on $A\bar{B}$ . The time sequence is

$$\bar{A}B_{(0)}AB_{(t_{A\to\bar{B}})}A\bar{B}_{(t_{A\to\bar{B}}+t_{\bar{B}\to A})}A\bar{B} \quad (11)$$

In order for this system change to a different state, external stimulation is needed.

For the system with double activation relationships, we have the stable solution

$$V(A \rightleftarrows B) = A_{t_{A\to B}}B_{t_{B\to A}}A + \bar{A}_{t_{\bar{A}\to\bar{B}}}\bar{B}_{t_{\bar{B}\to\bar{A}}}\bar{A} \quad (12)$$

If we consider a relationship with mixed activation and deactivation, i.e. $A \rightleftarrows B$, interesting results are obtained. The product of the values of the two edges is $(A_{t_{A\to B}}B + \bar{A}_{t_{\bar{A}\to\bar{B}}}\bar{B})(B_{t_{B\to\bar{A}}}A + \bar{B}_{t_{\bar{B}\to A}}A) = A_{t_{A\to B}}B_{t_{B\to\bar{A}}}\bar{A} + \bar{A}_{t_{\bar{A}\to\bar{B}}}\bar{B}_{t_{\bar{B}\to A}}A$ . In this situation, the starting and ending nodes of one item are the ending and starting nodes of the other. These two items can transfer back and forth from each other (*S4*). Then, the system oscillates. The value of the system is

$$V(A \rightleftarrows B) = A_{t_{A\to B}}B_{t_{B\to\bar{A}}}\bar{A}_{t_{\bar{A}\to\bar{B}}}\bar{B}_{t_{\bar{B}\to A}}A... \quad (13)$$

The components of the system oscillate in the order of $AB\bar{A}\bar{B}A$ with a period equal to $T = t_{A\to B} + t_{B\to\bar{A}} + t_{\bar{A}\to\bar{B}} + t_{\bar{B}\to A}$. This period parameter demonstrates the quantitative kinetic features of the present method. The active pulse width and inactive pulse width of A are $T_A = t_{A\to B} + t_{B\to\bar{A}}$ and $T_{\bar{A}} = t_{\bar{A}\to\bar{B}} + t_{\bar{B}\to A}$.

## 5.3 Merging operation.

A basic observation from the above six examples is that if two components have only an activation relationship (Equations 5, 9, 12), they will always have same states (both *ON* or *OFF*). But if only deactivation exists, the two components will always have different states, one takes *ON* and the other *OFF*. However, if the two components have both the activation and deactivation relationships, the system oscillates instead of being bistable. These observations can be used to simplify large systems. The procedure is referred to as merging operation that is consecutive nodes with activation interactions can be merged into one group with all its nodes having a same state. For example, *A*→*B*→*C*→*D* can be merged into *[ABCD]*, where *[]* is used to indicate that the nodes inside have activation relationships. This group of nodes is referred to as activation group node. The response times between nodes within the group can be calculated by adding all the response times between them. For examples

$$t_{A \to D} = t_{A \to B} + t_{B \to C} + t_{C \to D} \text{ and}$$

$$t_{\overline{A} \to \overline{D}} = t_{\overline{A} \to \overline{B}} + t_{\overline{B} \to \overline{C}} + t_{\overline{C} \to \overline{D}}$$

The above example can also be merged into *A→ [BCD]*, or other versions, depending on the need for specific problems. When there is a deactivation interaction among nodes, merging still applies if the states of the nodes are changed according to Equation 8. For example, a structure, *A → B –| C → D → E –| F –| G*, can be first merged into *[AB] –| [CDE] –| [F] –| [G]*. This relationship can be further merged into $[AB] \to \overline{[CDE]} \to [F] \to \overline{[G]}$. The response time can be calculated in a similar way, or

$$t_{A \to \overline{G}} = t_{A \to B} + t_{B \to \overline{C}} + t_{\overline{C} \to \overline{D}} + t_{\overline{D} \to \overline{E}} + t_{\overline{E} \to F} + t_{F \to \overline{G}} \text{ and}$$

$$t_{\overline{A} \to G} = t_{\overline{A} \to \overline{B}} + t_{\overline{B} \to C} + t_{C \to D} + t_{D \to E} + t_{E \to \overline{F}} + t_{\overline{F} \to G}.$$

### 5.4 Theorems for single variable systems

When systems have many nodes and each node has no more than one stimulator, the systems are called single variable networks. This type of networks can be trees or cycles. A cycle is a sequence of interacting components whose starting and ending members are the same (no root or leaf). Within a cycle, one can start from any component to travel back to that component again (closed loops). A tree is a sequence of interacting components, each of which can be passed only once when one travels from the starting component (root) to the ending ones (leafs). Linear structures and branched structures are trees. We have the three simple theorems for the single variable network systems.

**Theorem 1**: if a single variable network is a tree, the network has two states.

Proof: Applying the merging operation to all the consecutive nodes that have activation interactions, a single variable network can be transformed into new network that is composed of activation group nodes separated with deactivation edges. For example, *A → B –| C → D → E –| F –| G* becomes *[AB] –| [CDE] –| [F] –| [G]*. Using the terminology of graph theory, the merged sequence is a 2-color graph. That is the nodes or activation group node are either active or inactive (2 color or 2 types). The consecutive nodes are different in color from the ones next it.

If a 2-color graph is a tree (no cycle embedded inside), all the nodes can be further merged into two groups in the following way: assign number "1" to the starting node of the tree, "2" to the next, and continue until all the nodes have numbers. Because a tree does not have cycles embedded inside, the numbers can be assigned in such a way that any node with an even number is always next to nodes with odd numbers and vise versa. All the nodes with odd numbers can be in one group and all those with even numbers can be in the other group. Between these two groups, there are always deactivation interactions. Therefore, the color of each group has to be different. But inside the same group, there is no interaction among the activation group nodes. For example, the above structure becomes *{[AB], [F]} –|{[CDE], [G]}*. We call these two macro groups as “non-interactive macro groups” indicated with *{ }*. Similar to the situation of Equation 5, there are only two possible values for this merged structure. In the first value, all the nodes in one non-interactive macro group take active state, all the nodes in the other non-interactive macro group take inactive state. In a second value, the nodes take exactly opposite states to that in the first one.

**Theorem 2**: if a single variable network is a cycle, it has two stable states as long as the number of the deactivation relationships is an even number.

Proof: By applying the merging procedure described above, a cycle with even number of deactivation interactions can also be transferred into a graph with only two non-interactive macro groups. Similar to the situation of Equation 10, this structure also has two stable values.

**Theorem 3**: if a single variable network is a cycle and has odd number of deactivation relationships, the network does not have stable state. Instead, it oscillates with a characteristic period.

Proof: Considering a cycle with odd number of deactivation interactions (starting with 3. The case with 1 deactivation interaction was described with Equation 13), this type of systems can be merged into a structure with three non-interacting macro groups. For example,

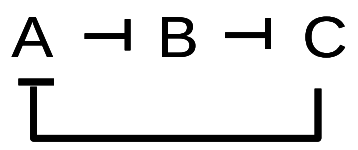

In graph theory, this cycle is a complete graph with three nodes. A complete graph is a graph in which all the nodes have common edges with all the other nodes. In order for all the nodes to have a different color from the ones next to it, a minimum of 3 colors are needed. Molecular networks have two colors (active and inactive). Using the same method as that used in Equation 13, it can be shown the color pattern oscillates in the order $A_{t_{A\to\bar{B}}}\bar{B}_{t_{\bar{B}\to C}}C_{t_{C\to\bar{A}}}\bar{A}_{t_{\bar{A}\to B}}B_{t_{B\to\bar{C}}}\bar{C}_{t_{\bar{C}\to A}}A\ldots$. This oscillation property is unique for a dynamic graph structure.

Theorem 1 suggests that if there is no cycle or closed loop, a network will respond to stimulations and then relax to its pre-stimulation state just like a simple two component system (as of the case of Equation 5). If there are more than two components in this type networks, the system may have delayed response to filter noise (see next Section 6). Theorems 2 and 3 explain why some systems are stable (homeostatic) and why other systems oscillate. More than this, these three theorems provided a simple way to determine the oscillation period or other kinetic properties of networks.

These three theorems have been mentioned in the literature [45, 46, and 66]. However, proofs of these theorems and discussions about their conditions have not been given previously. This is because previous methods such as numerical modeling and truth table listing [66] are excellent on studying individual cases. To proof these theorems, a general method is needed. The present dynamic logic representation provides such generality and lead to the verification of these theorems.

### 5.5 Multiple variable network systems

The simplest multiply variable network systems contain two stimulators and one responder. The two stimulators can be either activator or suppresser. They can have either *AND* or *OR* relationship. Therefore, there are 8 (= 2 X 2 X 2) cases. The first four cases include those in which the two stimulators are of the same types (both activators or suppressers. This is called coherent interactions in reference 46):

(1) $(A \wedge C) \rightarrow B$, *B* becomes active after both *A* and *C* become active.

(2) $(A \vee C) \rightarrow B$, *B* becomes active after either *A* or *C* becomes active.

(3) $(A \wedge C) -| B$, *B* becomes inactive after both *A* and *C* become active.

(4) $(A \vee C) -| B$, *B* becomes inactive after either *A* or *C* becomes active.

The other four cases include those in which the two stimulators are of different types (one is activator and the other is suppresser). This situation is called incoherent in reference 46. For example, $A \rightarrow B$ and $C -| B$. In this case, *B* becomes active after *A* is active and *C* is inactive. For convenience, these four cases can be presented in the following compact formats by using the relationship that $(C -| B)$ is equivalent to $(\overline{C} \rightarrow B)$ (Equation 8):

(5) $(A \wedge \overline{C}) \rightarrow B$ *B* becomes active after *A* becomes active and *C* becomes inactive.

(6) $(A \vee \overline{C}) \rightarrow B$ *B* becomes active after *A* becomes active or *C* becomes inactive.

(7) $(\overline{A} \wedge C) \rightarrow B$ *B* becomes active after *A* becomes inactive and *C* becomes active.

(8) $(\overline{A} \vee C) \rightarrow B$ *B* becomes active after *A* becomes inactive or *C* becomes active.

It can be seen that the case 7 (or 8) are essentially the same as the case 5 (or 6) because stimulators *A* and *C* can exchange. Therefore, there are 6 cases for the simplest multiple variable networks.

The values of the six cases can be calculated based on Equation 5. For example, let $D = A \wedge C$, the first case becomes $D \rightarrow B$. This is a simple activation system. The logic structure of the values of this case is $D_{t_{D \rightarrow B}} B + \overline{D}_{t_{\overline{D} \rightarrow \overline{B}}} \overline{B}$. However, $D$ and $\overline{D}$ actually have four different situations: $D = A \wedge C$ and $\overline{D} = \overline{A} \wedge C + A \wedge \overline{C} + \overline{A} \wedge \overline{C}$ [85]. For each situation, there is a separate response time. Then, we have

$$
\begin{aligned}
&V((A \wedge C) \rightarrow B) \\
&= (A \wedge C)_{t_{A \wedge C \rightarrow B}} B + (\overline{A} \wedge C)_{t_{\overline{A} \wedge C \rightarrow \overline{B}}} \overline{B} + (A \wedge \overline{C})_{t_{A \wedge \overline{C} \rightarrow \overline{B}}} \overline{B} + (\overline{A} \wedge \overline{C})_{t_{\overline{A} \wedge \overline{C} \rightarrow \overline{B}}} \overline{B}
\end{aligned}
\tag{14}
$$

The meaning of $(A \wedge C)_{t_{A \wedge C \rightarrow B}} B$ is that B will be activated after both *A* and *C* becomes active. The response time is $t_{A \wedge C \rightarrow B}$. The other items can be interpreted in a similar way. Those response times should be experimental determined parameters. The values of all the other cases can be calculated:

$$
\begin{aligned}
&V((A \vee C) \to B) \\
&= (A \wedge C)_{t_{A\wedge C \to B}} B + (\overline{A} \wedge C)_{t_{\overline{A}\wedge C \to B}} B + (A \wedge \overline{C})_{t_{A\wedge \overline{C} \to B}} B + (\overline{A} \wedge \overline{C})_{t_{\overline{A}\wedge \overline{C} \to \overline{B}}} \overline{B}
\end{aligned} \tag{15}
$$

$$
\begin{aligned}
&V((A \wedge C) -| B) \\
&= (A \wedge C)_{t_{A\wedge C \to \overline{B}}} \overline{B} + (\overline{A} \wedge C)_{t_{\overline{A}\wedge C \to B}} B + (A \wedge \overline{C})_{t_{A\wedge \overline{C} \to B}} B + (\overline{A} \wedge \overline{C})_{t_{\overline{A}\wedge \overline{C} \to B}} B
\end{aligned} \tag{16}
$$

$$
\begin{aligned}
&V((A \vee C) -| B) \\
&= (A \wedge C)_{t_{A\wedge C \to \overline{B}}} \overline{B} + (\overline{A} \wedge C)_{t_{\overline{A}\wedge C \to \hat{B}}} \overline{B} + (A \wedge \overline{C})_{t_{A\wedge \overline{C} \to \overline{B}}} \overline{B} + (\overline{A} \wedge \overline{C})_{t_{\overline{A}\wedge \overline{C} \to B}} B
\end{aligned} \tag{17}
$$

$$
\begin{aligned}
&V((A \wedge \overline{C}) \to B) \\
&= (A \wedge \overline{C})_{t_{A\wedge \overline{C} \to B}} B + (\overline{A} \wedge C)_{t_{\overline{A}\wedge C \to \overline{B}}} \overline{B} + (A \wedge C)_{t_{A\wedge C \to \overline{B}}} \overline{B} + (\overline{A} \wedge \overline{C})_{t_{\overline{A}\wedge \overline{C} \to \overline{B}}} \overline{B}
\end{aligned} \tag{18}
$$

$$
\begin{aligned}
&V((A \vee \overline{C}) \to B) \\
&= (A \wedge \overline{C})_{t_{A\wedge \overline{C} \to B}} B + (\overline{A} \wedge C)_{t_{\overline{A}\wedge C \to \overline{B}}} \overline{B} + (A \wedge C)_{t_{A\wedge C \to B}} B + (\overline{A} \wedge \overline{C})_{t_{\overline{A}\wedge \overline{C} \to B}} B
\end{aligned} \tag{19}
$$

Network nodes with more stimulators can be analyzed in similar way. For example, a node has three stimulators, $(A \wedge C) \vee D \to B$, can be simplified into $E \vee D \to B$, where $E = (A \wedge C)$. The logic expression of $E \vee D \to B$ (Equation 15) is

$$
\begin{aligned}
&V((E \vee D) \to B) \\
&= (E \wedge D)_{t_{E\wedge D \to B}} B + (\overline{E} \wedge D)_{t_{\overline{E}\wedge D \to B}} B + (E \wedge \overline{D})_{t_{E\wedge \overline{D} \to B}} B + (\overline{E} \wedge \overline{D})_{t_{\overline{E}\wedge \overline{D} \to \overline{B}}} \overline{B}
\end{aligned}
$$

Substituting $E = (A \wedge C)$ and $\overline{E} = (\overline{A} \wedge \overline{C}) + (\overline{A} \wedge C) + (A + \overline{C})$, we have

$$
\begin{aligned}
&V(((A \wedge C) \vee D) \to B) \\
&= (A \wedge C \wedge D)_{t_{A\wedge C\wedge D \to B}} B + (\overline{A} \wedge \overline{C} \wedge D)_{t_{\overline{A}\wedge \overline{C}\wedge D \to B}} B + (\overline{A} \wedge C \wedge D)_{t_{\overline{A}\wedge C\wedge D \to B}} B \\
&+ (A \wedge \overline{C} \wedge D)_{t_{A\wedge \overline{C}\wedge D \to B}} B + (A \wedge C \wedge \overline{D})_{t_{A\wedge C\wedge \overline{D} \to B}} B + (\overline{A} \wedge \overline{C} \wedge \overline{D})_{t_{\overline{A}\wedge \overline{C}\wedge \overline{D} \to \overline{B}}} \overline{B} \\
&+ (\overline{A} \wedge C \wedge \overline{D})_{t_{\overline{A}\wedge C\wedge \overline{D} \to \overline{B}}} \overline{B} + (A \wedge \overline{C} \wedge \overline{D})_{t_{A\wedge \overline{C}\wedge \overline{D} \to \overline{B}}} \overline{B}
\end{aligned} \tag{20}
$$

## 6. SIGNAL TRAVELING IN NETWROKS

### 6.1 Signal travels in linear chains

A linear chain has $n$ nodes, $N_1$, $N_2$, …$N_n$, each of which stimulates the one next (right) to it. Only $N_1$ accepts and responds to an external signal ($E$). The scheme of this network is

$E \rightarrow N_1 \rightarrow N_2 \rightarrow \ldots \rightarrow N_{n-1} \rightarrow N_n$

Before time $t_0$, $N_1$ is inactive. At $t_0$, $E$ activates $N_1$. This activation signal propagates into the chain. Node $N_i$ responds to the signal after a time

$$t_{N_1 \to N_i} = \sum_{j=1}^{i-1} t_{N_j \to N_{j+1}} \tag{21}$$

If the external signal turns $N_1$ off at time $T$, $N_i$ responds to the signal after a time

$$t_{\overline{N_1} \to \overline{N_i}} = \sum_{j=1}^{i-1} t_{\overline{N_j} \to \overline{N_{j+1}}} \tag{22}$$

If the external signal is a pulse, each node has a pulse responding to the external signal. The pulse width of node $N_i$ is

$$T_i = T + \sum_{j=1}^{i-1} (t_{\overline{N_j} \to \overline{N_{j+1}}} - t_{N_j \to N_{j+1}}) \tag{23}$$

This simply indicates that the pulse width of each node is different. Each node will change the pulse width by an amount of $(t_{\overline{N_j} \to \overline{N_{j+1}}} - t_{N_j \to N_{j+1}})$ that can be either an increase or decrease. If the pulse width is decreased as the signal travels through the chain, the signal is terminated when the active pulse width of a node gets shorter than the activation response time needed for this node to activate the next one. Thus, for a same chemical signal, if it has different pulse width ($T$), it travels to a different depth or position of the chain and results in different biological responses. If the pulse width is increased as the signal travels, the nodes of the chain can have broader active pulse width than the external signal. Then a short external stimulation can result in an elongated response. Therefore, a same type of chemical signal can stimulate more than one response in chains that have different abilities to modify pulse width. Since pulse width is related to the frequency of signal, the signal with different frequency may stimulate different biological reactions.

The time sequence of each node can be calculated as

$$(t_{N_1 \to N_i})\, N_i\, (T + t_{\overline{N_1} \to \overline{N_i}})\, \overline{N_i} \tag{24}$$

Activation relationships were assumed in this situation. But networks with deactivation relationships can be treated in the same way.

### 6.2 Signals travel in bistable loops

A biostable loop is composed of $n$ nodes with the following relationship:

$E \rightarrow N_1 \rightarrow N_2 \rightarrow \cdots \rightarrow N_{n-1} \rightarrow N_n$ (with a feedback arrow from $N_n$ back to $N_1$)

Where, $N_{i-1}$ activates $N_i$. $E$ is an external signal that works together with $N_n$ to activate $N_1$, for example, $E \vee N_n \to N_1$. Without $E$, the system is a bistable cycle as there is no deactivation interaction (Theorem 2). Let's assume $E$ have $OR$ relationship with $N_n$. When $E$ is inactive, it has no effect on the system but when it is active, it always activates $N_1$ (Switch). Assume $E$ and all the nodes of the system are inactive at beginning. At time $t_0$, $E$ become active and starts to activate $N_1$. After $t_{E \to N_1}$, $N_1$ becomes active. This active signal travels into the cycle and it returns to $N_1$after a time $t_{N_1 \to N_1}$.

$$t_{N_1 \to N_1} = \sum_{j=1}^{n-1} t_{N_j \to N_{j+1}} + t_{E \wedge N_n \to N_1} \tag{25}$$

After then, the entire network is activated. However, if $E$ is a pulse with an active pulse width $T_E < t_{N_1 \to N_1}$, that is $E$ become inactive before $N_1$ is activated by $N_n$, can the entire network still be activated? The answer is it is possible. After $E$ becomes inactive, it has no effect on the system. $N_1$ is deactivated by $N_n$ after $t_{\overline{N}_n \to \overline{N}_1}$. Then this pulse propagates into the cycle. As discussed in Section 6.1, each node can change the pulse width by an amount of $t_{\overline{N_j} \to \overline{N_{j+1}}} - t_{N_j \to N_{j+1}}$. There are two cases. If the pulse width decreases as traveling in the cycle, at a certain node where the width is shorter than the time needed for that node to active the next one, the pulse signal terminates. If the pulse width increases as traveling through the cycle, after traveling through the cycle for a certain number of turns, the entire cycle become active. This example simply suggests that different cycles react to a same external pulse signal differently. The cycle can be either entirely activated or be activated for a short period of time and return to inactive state. Systems where $N_n$ and $E$ have $AND$ relationship can be analyzed in a similar way.

### 6.3 Signal travel in oscillation loops.

Signal traveling in oscillation loop can be analyzed in the same way as that for bistable loops. The difference is that the states of the nodes of oscillation loops change between active and inactive alternatively while it stays the same in a bistable cycle.

### 6.4 Single variable systems are either stable or oscillating

Systems with their nodes having no more than one stimulator are either stable or oscillating. This can be proved as the follows.

Prove: a single variable system can have only two types of structures. One is a tree without any cycle (Figure 2A) and the other is a structure with only one cycle embedded inside (Figure 2B). For the latter case, if the cycle has branches, the nodes within the cycle are activators (node *D* in Figure 2B). Otherwise, the nodes in the cycles would have more than one activator (node *D* in Figure 2C). If a system has two or more cycles, there is at least one node in the system has more than one activator (node *E* in Figure 2D).

If a single variable system does not have a cycle, the system is a tree. According to Theorem 1, the system has two states. If a system has a cycle and there are even numbers of deactivation interactions inside the cycle, the system has two states as well (Theorem 2). If a system has a cycle that has odd numbers of deactivation interactions, the system oscillates (Theorem 3).

Since either stable states or oscillating states are well defined, it can be deduced that single variable networks do not have chaotic behaviors.

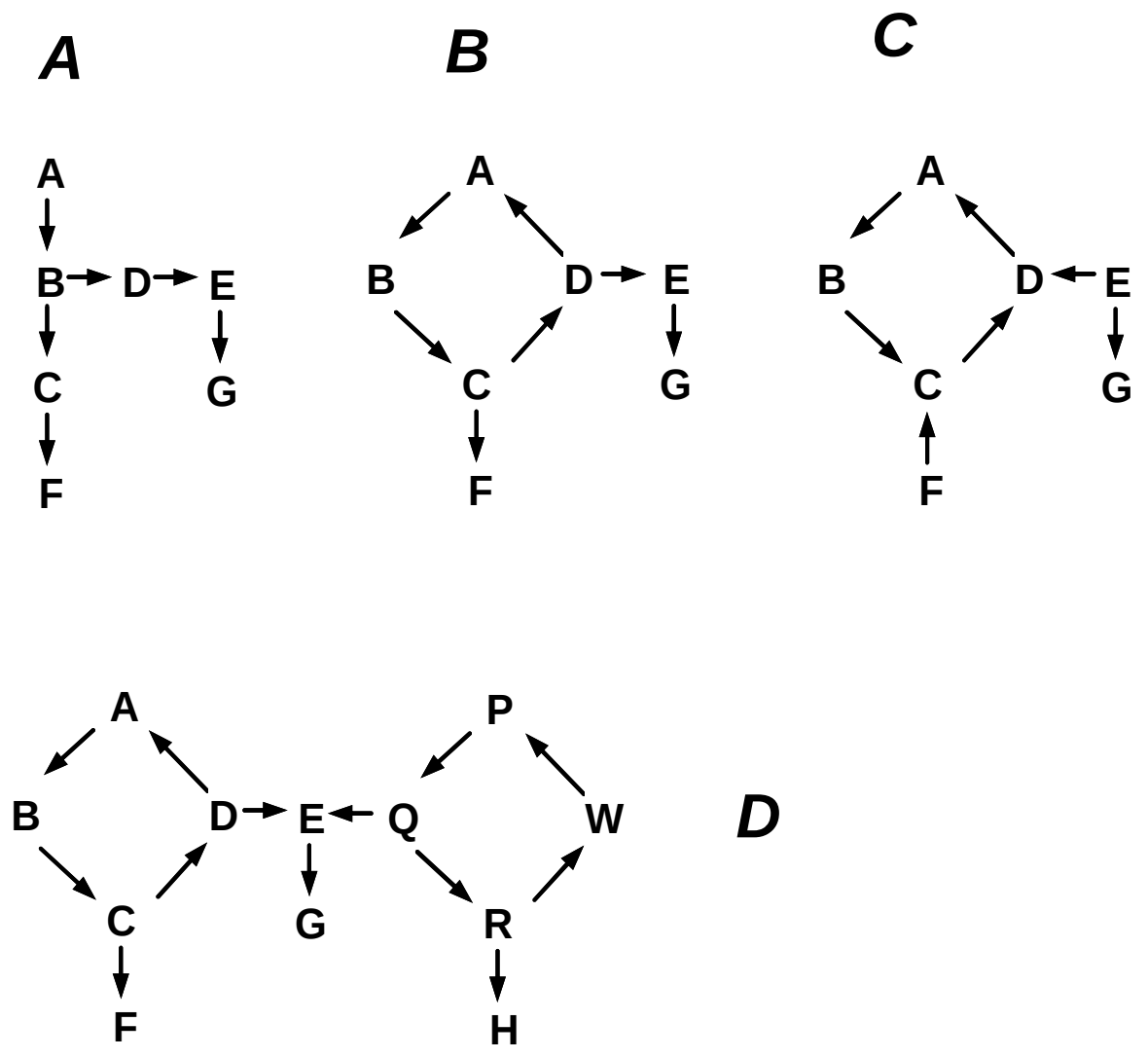

Figure 2. Single variable networks have two types of structures, *A* and *B*. Structure *C* and *D* are multiple variable networks.

## 7. GENERAL PROCEDURE

A system that has more nodes may be analyzed by using multiplication procedure as discussed above (Procedures S1 to S4). However, calculation can be very complicated if nodes have many stimulators and responders. Here, we introduce a general method to calculate the values of any multiple variable system. The target of the calculation is to determine the time sequences of the system based on the observed interactions among its nodes. This general procedure is analogous to numerical methods for ordinary differential equations (ODE).

For a system composed of *n* nodes, $N_1$ to $N_n$, first step is to identify state conditions of all the nodes based on the available experimental observations. For example, if it is observed that $N_l$ and $N_m$ jointly activate (*AND*) $N_j$, the active and inactive condition of $N_j$ are

$$t_{N_l \wedge N_m \to N_j} \, N_j \qquad (26)$$

$$t_{\overline{N}_l \wedge N_m \to \overline{N}_j} \, \overline{N}_j + t_{N_l \wedge \overline{N}_m \to \overline{N}_j} \, \overline{N}_j + t_{\overline{N}_l \wedge \overline{N}_m \to \overline{N}_j} \, \overline{N}_j \qquad (27)$$

Where “+” in Equation 27 indicates that there are three deactivation conditions, each of them can lead to $N_j$ to become inactive. Conditions of all the nodes can be determined in the same way. This is analogous to define the numerical equations for ODEs.

A second step is to choose initial states for all the nodes and use them as the initial point for a time sequence of the system (initial condition).

A third step is to determine the time sequence of the system based on the initial condition and the state conditions (Equations 26 and 27) of all the nodes. This is done by increasing time by a small step $\delta t$ (< all the response times), checking all the nodes and determining whether their state conditions are met so they can be activated or deactivated. If any node is activated (deactivated), append the time and new state of the node to the time sequence. Update the node states and continue this operation until the sequence become stable or oscillating over time. This stable or oscillating sequence is a value of the system.

A fourth step is to repeat third step with different initial conditions to find out all the other values of the system.

For systems that have more than one value, the initial sequence in Step 4 can be chosen in such a way that it is as different as possible from the values of the system that have been obtained prior to it. This is similar to numerical modeling where, in order to find different steady state values, the initial values are selected to be far from each other.

## 8. APPLICATIONS IN BIOLOGICAL NETWORKS

The results based on the present method can immediately explain the behaviors of a number of stable molecular networks [22, 24, 27, 42-45, 63] and a number of oscillatory networks reported in literature such as the synthetic oscillator [34, 42, 45], lac operon [42], cAMP oscillation of Dictyostelium cells [86-88], and cell cycles [35, 36, 39, 83, 89-90]. Theorems 2 and 3 also explain the observations made by Ferrell et al. [22, 34, 43] and Thomas [66] that cycles with double deactivation interactions were bistable and cycles with odd number of deactivation interactions would oscillate. More than explaining the literature results, kinetic parameters such as oscillation periods can be readily determined based on the present biologic method. Here we give a few examples to demonstrate how to use the present method.

### 8.1 Bistable network: the MAPK cascade in frog oocytes

The mos-mitogen-activated protein kinase (MAPK) kinase (MEK)-p42 MAPK cascade in Xenopus oocytes contains the following protein interaction structure (43):

Mos → MEK → p42 MAPK → Rsk ⊣ Myt1 ⊣ Cdc2/Cyclin B

What steady state behaviors can this network have?

**Solution:**

Using the merging operation we can convert the original structure into,

$$[Cdc2/CyclinB, Mos, MEK, P42MAPK, Rsk] \quad -| \quad Myt1$$

This new structure contains two deactivation edges. Based on Theorem 2, this network should have two stable solutions:

$$\overline{Cdc2/CyclinB}, \overline{Mos}, \overline{MEK}, \overline{P42MAPK}, \overline{Rsk}, Myt1 \qquad (29A)$$

$$Cdc2, CyclinB, Mos, MEK, P42MAPK, Rsk, \overline{Myt1} \qquad (29B)$$

An external signal can turn system form one state to the other. There might be variation in response times from cell to cell. The state to state transition of different cells may occur at different times.

Ferrell et al. reported, when oocytes were stimulated with progesterone, individual cells were either nearly completely activated or not activated. The activation time differs from cell to cell. There was no oocyte with intermediate activation [43]. Ferrell et al. attributed this behavior to the structures of the network. Here, we demonstrated this structure-behavior relationship with the dynamic logic method.

## 8.2 Feed-forward Loop (FFL)

Feed-forward loop is one of most abundant motifs. Reference 46 has a comprehensive description about this type of motif. One of the most common FFL structures is described in Figure 3. It contains three nodes, *X, Y,* and *Z*. *X* activates *Y* then *X* and *Y* jointly activate *Z*. The relationship between *X* and *Y* can be either *AND* or *OR*. The final result is that *Z* is activated or deactivated by *X*. People may ask why biological systems do not simply us $X \rightarrow Z$ instead of the FFL motif with an extra component *Y*? This is because this motif provides possibility to manage schedules of biological events in various ways such as first in-first out (*FIFO*) and just-as-needed compared to the simple structure $X \rightarrow Z$ that often has first in-first out schedule (*FIFO*) [46]. Reference 46 presents an excellent descriptive analysis about *E. Coli* cells that have 6 *Z* operons controlled by a same pair of *X* and *Y* in *OR* relationship. This FFL structure provides possibility for the E. Coli cells to control their flagellar motor assembly process in a just-as-needed schedule. Here, we can use the above logic method to re-demonstrate this possibility in a simpler way.

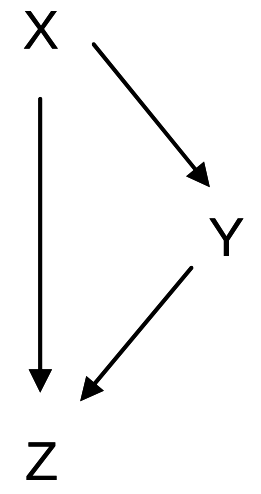

Figure 3. One example of feed-forward loops (FFL). *X* activates *Y*. *X* and *Y* together (*OR*) activate *Z*.

**Solution:**

We first analyze a system that uses an *X* to control multiple *Z*s through $X \to Z_i$ edges. The time sequence of $X \to Z_i$ is $(0)\overline{Z_i}\,{}_{(t_{X\to Z_i})}Z_i\,{}_{(T+t_{\overline{X}\to\overline{Z_i}})}\overline{Z}_i$ (*X* is activated at instant 0 by an external signal and this external signal turns *X* off at instant *T*). According to reference 46, activation processes usually are based on concentrations of *X*. Then, if $t_{X\to Z_i} > t_{X\to Z_j}$, we would have $t_{\overline{X}\to\overline{Z_i}} < t_{\overline{X}\to\overline{Z_j}}$. This means the operon that is activated first will is deactivated last (first in-last out).

For a system that is composted of the above FFL,

$$X \to Y$$

$$X \vee Y \to Z_i$$

we can calculate its time sequence as the follows. The time sequence of $X \to Y$ is

$$\overline{X}\,\overline{Y}_{(0)}\,X\overline{Y}_{(t_{X\to Y})}\,XY_{(T)}\,\overline{X}Y_{(T+t_{\overline{X}\to\overline{Y}})}\,\overline{X}\,\overline{Y} \tag{30}$$

Comparing this with Equations 3 and 15, we have the following time sequence for operon $Z_i$,

$$(0)\overline{Z_i}\,{}_{\min(t_{X\to Z_i},\ t_{X\to Y}+t_{Y\to Z_i})}Z_i\,{}_{(T+t_{\overline{X}\to\overline{Y}}+t_{\overline{X}\wedge\overline{Y}\to\overline{Z_i}})}\overline{Z}_i \tag{31}$$

The function *min* of the above equation gives the shorter response time of the two possible processes, one is *X* directly activates $Z_i$ and the other is *X* activates *Y* and *Y* activates $Z_i$. Compared to the time sequence of the system $X \to Z_i$, one can see that the turn-off of $Z_i$ is controlled by not only *X* but also *Y*. The interactions between *Y* and $Z_i$ is independent of the interactions between *X* and $Z_i$. Then, this extra dimension provides possibilities that $t_{\overline{X}\wedge\overline{Y}\to\overline{Z_i}} > t_{\overline{X}\wedge\overline{Y}\to\overline{Z_j}}$ even if $t_{X\to Z_i} > t_{X\to Z_j}$, or vise versa. Thus, the turn-on/off of $Z_i$ does not have to follow the first in-last out schedule as that for $X \to Z_i$. It can be first in-first out (FIFO) or just-as-needed. *E.Coli* cells have six operons ($Z_1$ to $Z_6$) for their flagellar motor assembly. This function allows the cells to control their $Z_i$ to be turned on/off in a just-as-needed manner [46].

*X* and *Y* can have *AND* relationship as well. In this case, the time sequence of *Z* is,

$$\overline{Z}_i\,{}_{(t_{X\to Y}+t_{X\wedge Y\to Z_i})}Z_i\,{}_{(T+\min(t_{\overline{X}\to\overline{Z_i}},\ t_{\overline{X}\to\overline{Y}}+t_{\overline{Y}\to\overline{Z_i}}))}\overline{Z}_i \tag{32}$$

Compared this sequence to that of $X \rightarrow Z_i$, one can see the activation of $Z_i$ is delayed. This delay mechanism may function as a noise filter. Any signal shorter than $t_{X \rightarrow Y} + t_{X \wedge Y \rightarrow Z_i}$ will be ignored.

### 8.3 Budding yeast mitosis

Budding yeast cell mitosis circle is one of the most studied systems [83, 89-90]. A number of proteins are involved the cycle. Li et al. simplified the system into 11 nodes. The cell cycle was simulated by solving a group of dynamic equations with *G1* phase as the initial value. It was found the systems evolved from *G1, S, G2,* through *M* and back to *G1* phases. If delete some edges from the system, for a significant amount cases the system was still able to go through the right cell cycle [83]. It was then concluded that the cell cycle was inherently robust. However, because quantitative response times were not available, it is not clear whether the robustness is also true against variations of response times. Here we will use the present dynamic logic method to demonstrate that this network is also robust against variations in response times.

**Solution:**

We adopted the logic structures of cell cycle presented in references 83 and 89-90 (Figure 4 for graph and Appendix for logic equations). The analysis was done with Mathematica® as the follows.

Because there is no response time reported, we set all of them to be a same number, *t* (randomly set to 1). Then we determined the state conditions of all the nodes based on the logic relationships (as listed in Appendix). For example, node *SBF* is controlled by nodes *Cln3* and *Clb1,2* through a logic relationship, $C\ln 3 \wedge \overline{Clb1,2} \rightarrow SBF$. The state conditions of *SBF* are

$$t_{C\ln 3 \wedge \overline{Clb1,2} \rightarrow SBF}\ SBF \quad (33A)$$

$$t_{\overline{C\ln 3} \wedge \overline{Clb1,2} \rightarrow \overline{SBF}}\ \overline{SBF} \quad (33B)$$

$$t_{Clb1,2 \wedge Clb1,2 \rightarrow \overline{SBF}}\ \overline{SBF} \quad (33C)$$

$$t_{\overline{Clb1,2} \wedge Clb1,2 \rightarrow \overline{SBF}}\ \overline{SBF} \quad (33D)$$

In a similar way, the state conditions of all the other nodes can be determined. For simplicity, they are not listed here. Each node has two values. One is its current state and the other is the time at which it made its most recent state change.

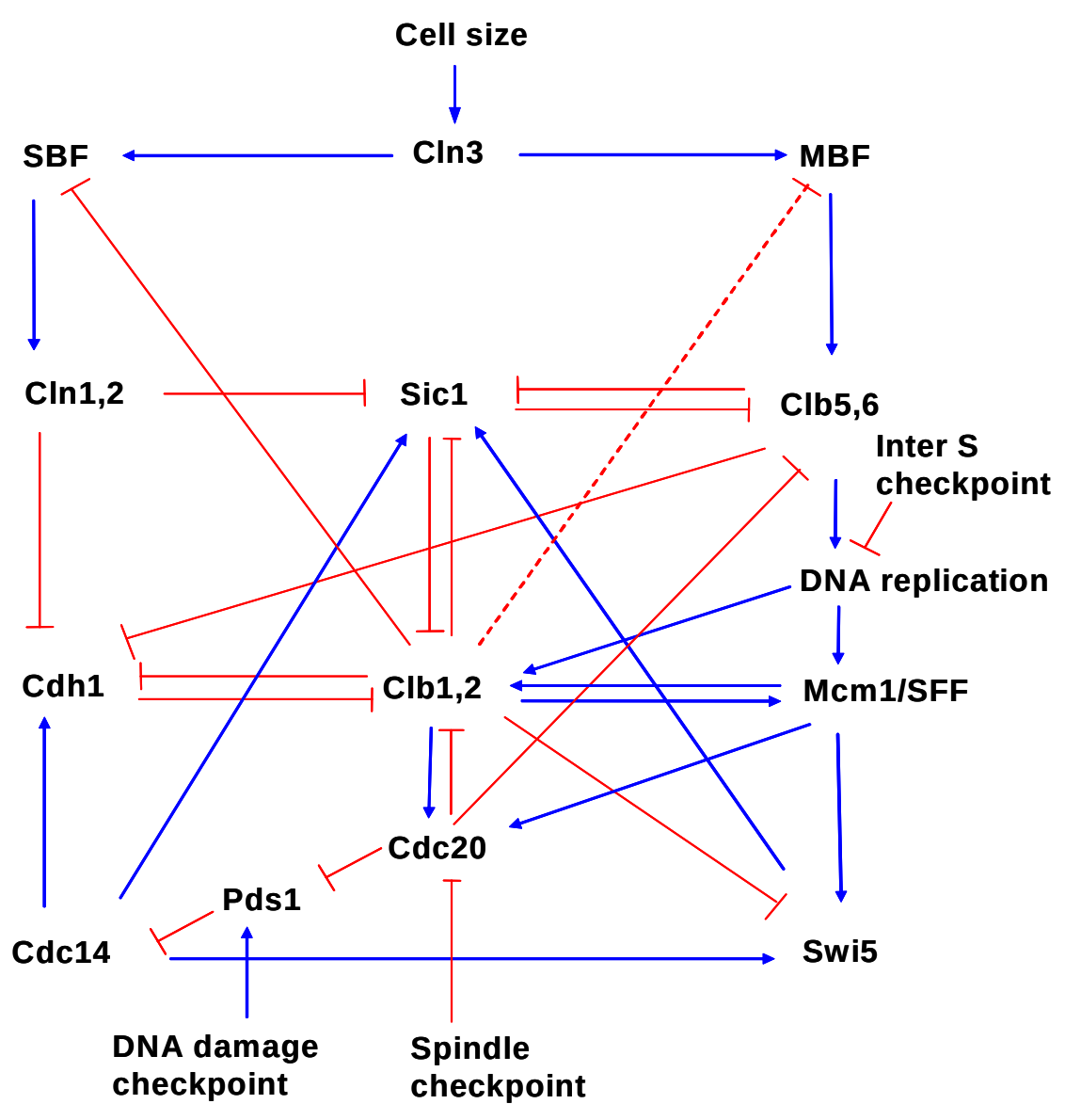

Figure 4. Simplified molecular network of yeast cell mitosis cycle adapted from reference 83 and 89-90.

In a second step, we set all the nodes to be active except that the three checkpoints were set to be inactive. We used this set as an initial value for a time sequence of the system. Set time to be zero.

A third step was to determine the evolution of the time sequence of the system based on the initial condition (all nodes being active) and the state conditions of all the nodes (state conditions of each node is similar to 33A – 33D). This was done by increasing time by a step $\delta t$ (e.g. 0.005), checking all the nodes and determining whether their state conditions were met so they could be activated or deactivated. If any node was activated (or deactivated), update the state and state change time of that node and append that node to the time sequence of the system. This process was repeated until the sequence became oscillating over time. This oscillating sequence is show in Figure 5. As expected, the sequence represents a change from *G1*, to *S*, to *G2*, and to *M*. This is the same order of cell mitosis cycle [83]. In order to check whether there is other solution, we reset the initial value by let every node be inactive (except node *cln3* that was kept active because this node receives external stimulation for cell cycle to start). It was found the system still reached the right cycle from *G1* to *M*. The initial values for this second calculation was very different from that of the first calculation where most of the nodes were set to be active.

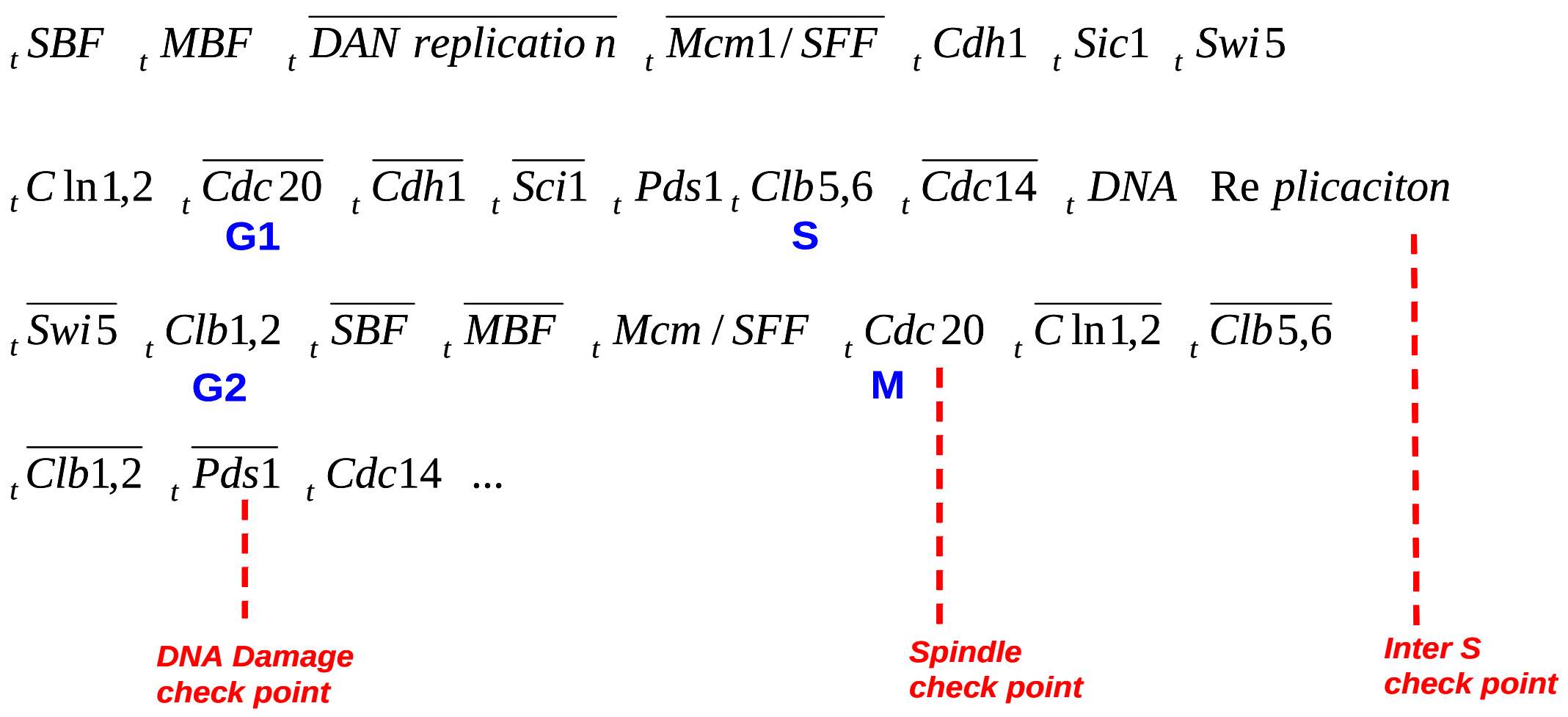

Figure 5. Calculated mitosis cycle of yeast cells. Due to response times are not available and all of them are simply set to *t*.

Then we changed the response times one by one and calculated the time sequence of the system. We observed that changes in most of the response times by two to three orders of magnitude (*0.01* to *10*) did not disturb the cell cycle (*G1* through *M*). A few response times had narrower ranges, such as the activation times of nodes C*dc20* and S*wi5* and deactivation times of the nodes *Mcm/SFF*, *Pds1*, and *Cdc20*. The change in these response times had to be less than one order of magnitude in order for the cell cycle to be unchanged. This new observation suggests that the mitosis cycle of yeast cells is robust against variation of most response times.

We also observed that disconnecting some edges did not disturb this cell cycle. However, the oscillation stopped if disconnecting the edge from *DNA Replication* to *Clb1,2*. This observed sensitivity is consistent with the experimental observations: cell does not move into the next phase if its *NDA replication* is not completed. Also, it was observed the cycle was stopped if any of the three check points (Figure 4) was on.

## 9. DISCUSSIONS

The application of logic method in biological networks has been promoted by Kauffman [64] and Thomas et al [66]. The method is based on truth tables. The present method is based on two types of sequences. One is the dynamic logic sequences that characterize the interactions among nodes and the other time sequences that represent dynamics of networks. Both Thomas' and the present methods can be used to determine the steady state behaviors of molecular networks with minimal kinetic parameter inputs. Therefore, they are expected to be useful alternatives to the quantitative modeling methods. However, the present method has some advantages over other logic methods and quantitative modeling methods.

First, the existing truth table methods did not incorporate response times (kinetic properties of the networks) of nodes into analysis. The logic kinetic sequences and time sequences of the present method naturally combine the logic process and kinetic parameters together. Therefore, both the steady state properties and kinetic properties can be studied.

Secondly, truth table method is a case-by-case approach. It is not designed for general analysis. Also, it can be complicated as systems get large. The present dynamic logic method can be used to study general properties of networks. Because of this, the three theorems describing the steady state properties of single variable networks have been approved for first time in the present paper although these three theorems were discovered previously with truth table methods.

Thirdly, the present method can be used to analyze network at different levels of details depending on the availability of quantitative kinetic parameters. If a complicated network system can be decomposed into modules and motifs, these modules or motifs may be readily studied with the above methods (as examples 8.1 and 8.2). Secondly, by using the general procedure, networks can be analyzed without quantitative kinetic parameters. This was demonstrated in the yeast cell mitosis example. Interestingly, even without reliable parameters, the analysis provided correct oscillation properties as well as the robustness property and sensitivity of network structure. Thirdly, if kinetic parameters are available, more quantitative analysis can be expected. Oscillation periods and other quantitative characteristics of network dynamics could be calculated.

These advantages may help to fill in the gap between the need and availability of quantitative parameters of biological interactions. Molecular biology is excellent at qualitatively observing [48]. Quantitative kinetic parameters are often not reported. Kinetic methods, either deterministic or stochastic, require parameters to quantify themselves. The present method provides an opportunity for biologists to identify steady state behaviors of networks quickly and directly without using reliable parameters.

The response time in the current method quantitatively measures the integral kinetics of the molecular interactions (the rate constant measures the differential kinetics). It answers the question of how long it takes for a biological system to respond to stimulation. Thus, it inherently contains the information about sensitivity, duration, and the relationships of interactions. However, it does not give information about the transient process of the response. Whether this is quantitative enough depends on what level of detail one likes to focuses on. In terms of experimental measurements, less data points are needed to determine response times compared to the determination of a full transient process. Therefore, the response time may be a realistic parameter to quantify biology networks.

While the simplicity is an advantage of using the present method, it can be a drawback as well. Simplicity of the present method comes from the discrete state assumption (actually binary states). Although there has been an increasing amount of evidence to support this assumption, systems in which nodes take continuous values or multiple discrete values may exist. For these systems, the bistable or oscillating behaviors predicted by the present method might not apply. Multiple discrete state approach is needed fro those systems.

**ACKNOWLEDGEMENTS**

I am in deep debt to Darrel Untereker for his critical reviewing and editing the paper.

**APPENDIX**

Yeast cell motisis logic was adopted from the reference 83, 89-90 and are summarized as the follows

$$C\ln 3 \wedge \overline{Clb1,2} \rightarrow SBF$$

$$C\ln 3 \wedge \overline{Clb1,2} \rightarrow MBF$$

$$SBF \rightarrow C\ln 1,2$$

$$(C\ln 1,2 \vee Clb5,6 \vee Clb1,2) \wedge \overline{Swi5 \vee Cdc14} \rightarrow \overline{Sic1}$$

$$(\overline{C\ln 1,2} \wedge \overline{Clb1,2} \wedge \overline{Clb5,6}) \wedge Cdc14 \rightarrow Cdh1$$

$$\overline{Sic1} \wedge MBF \wedge \overline{Cdc20} \rightarrow Clb5,6$$

$$Clb5,6 \wedge \overline{S\ \ check\ \ po\text{int}} \rightarrow DNA\ \ replication$$

$$DNA\ \ replication \vee Clb1,2 \rightarrow Mcm1/SFF$$

$$(DNA\ \ replicatio\ n \vee Mcm1/SFF) \wedge \overline{Cdh1} \wedge \overline{Sic1} \wedge \overline{Cdc20} \rightarrow Clb1,2$$

$$Clb1,2 \wedge Mcm1/SFF \wedge \overline{Spindle\ \ Check\ \ po\text{int}} \rightarrow Cdc20$$

$$Cdc20 \wedge \overline{DNA\ \ damage\ \ Check\ \ po\text{int}} \rightarrow \overline{Pds1}$$

$$Pds1 \rightarrow \overline{Cdc14}$$

$$(Cdc14 \vee Mcm1/SFF) \wedge \overline{Clb1,2} \rightarrow Swi5$$